\documentclass[a4paper,12pt,epsfig,superscriptaddress,showpacs]{article}
\usepackage{cite} 
\usepackage[utf8]{inputenc}
\usepackage{graphicx}
\usepackage{amsmath}
\usepackage{amssymb}
\usepackage{cmap}
\usepackage{bm}
\usepackage{array}
\usepackage{epsfig}
\usepackage{array}
\usepackage{color}
\usepackage{cite}
\usepackage{latexsym}
\usepackage{amscd, xypic}

\usepackage[left=1.0in, right=1.0in, top=1.0in, bottom=1.0in]{geometry}

\title{\bf \huge{Colormagnetic confinement in the  quark-gluon thermodynamics}\bigskip } 

\author{
M.S.~Lukashov$^{a,b,}$\thanks{lukashov@phystech.edu}\,\, and Yu.A.~Simonov$^{a,}$\thanks{simonov@itep.ru}\, 
\\ \\
$^a$ \small{\em Alikhanov Institute for Theoretical and Experimental Physics,}\\
\small{\em Moscow 117218, Russia}\\
$^b$ \small{\em Moscow Institute of Physics and Technology,}\\
\small{\em Dolgoprudny 141700, Moscow Region, Russia}\\
}\bigskip
\date{\today}

\newcommand{\be}{\begin{equation}}
\newcommand{\ee}{\end{equation}}

\def\la{\mathrel{\mathpalette\fun <}}

\def\fun#1#2{\lower3.6pt\vbox{\baselineskip0pt\lineskip.9pt
\ialign{$\mathsurround=0pt#1\hfil ##\hfil$\crcr#2\crcr\sim\crcr}}}

\newcommand{{\SD}}{\rm SD}

\newcommand{\vex}{\mbox{\boldmath${\rm x}$}}

\newcommand{\veu}{\mbox{\boldmath${\rm u}$}}

\newcommand{{\Mc}}{\mathcal{M}}

\newcommand{\lan}{\langle}
\newcommand{\ran}{\rangle}

\begin{document}

\maketitle
\begin{abstract}
Nonperturbative effects in the  quark-gluon thermodynamics are studied in the
framework   of Vacuum Correlator Method. It is shown, that two  correlators:
colorelectric $D_1^E(x)$ and colormagnetic $D^H(x)$, provide  the Polyakov line
  and
the colormagnetic confinement in the spatial planes respectively. As a result
both effects produce the realistic behavior of $P(T)$ and $s(T)$, being in good
agreement with numerical lattice data.
\end{abstract}

\newpage

\section{Introduction }
The idea of a new phase of the QCD matter above some critical temperature has
appeared soon after the discovery of QCD, namely in \cite{1,2,3} were
formulated the first principles of weak interacting quark-gluon medium, named
the quark-gluon plasma (QGP).

The first lattice studies \cite{4,5,6} have supported this idea, and it was
soon realized that high-energy ion collisions can be used to create QGP, see
\cite{7} for a recent review and references.

The subsequent  lattice studies of QGP and thermal transitions   has discovered
a variety of  sudden and complicated features of the QGP behavior, especially
near the transition temperature $T_c$ \cite{8}. At present the high accuracy
lattice data are obtained for $n_f =2+1$ QCD in the wide   temperature region
\cite{37,39, 40}.

 An important
progress was made at large $T$ in the  framework of the perturbation theory
(Hard Thermal Loop (HTL) theory) \cite{15*}, where terms up  to $O(g^6)$ have
been taken into account.

However, in the  region  150 MeV$< T<$ 600 MeV the nonperturbative (np) effects
are most important, which can be taken into account in the framework of the
Vacuum Correlator Method  (VCM), to be used below.

This method  was suggested at the   end of 80's in \cite{16,17}, stating, that
the basic origin of  the nonperturbative  dynamics in QCD at zero or nonzero
$T$ is connected with the vacuum  gluonic fields, appearing in the form of
gluon vacuum correlators.  In FCM the confinement  follows from nonzero
quadratic correlator  $D^E (x-y)$ of colorelectric (CE) fields $E_i^a (x)$,
which produce scalar linear confining interaction  $V_D^{(\rm lin)} (r)$, while
correlators $D^H (x-y)$ of colormagnetic (CM) fields $H_i^a (x)$, are
responsible for confinement in spatial  surfaces.


$$ \frac{g^2}{N_c}\lan\lan {\rm
Tr} E_i(x)\Phi
E_j(y)\Phi^\dagger\ran\ran=\delta_{ij}\left(D^E(u)+D_1^E(u)+u^2_4\frac{\partial
D_1^E}{\partial u^2}\right)+ u_iu_j\frac{\partial D_1^E}{\partial u^2},$$ \be
\frac{g^2}{N_c}\lan\lan {\rm Tr} H_i(x)\Phi
H_j(y)\Phi^\dagger\ran\ran=\delta_{ij}\left(D^H(u)+D_1^H(u)+\veu^2\frac{\partial
D_1^H}{\partial\veu^2}\right)- u_iu_j\frac{\partial D_1^H}{\partial
u^2},\label{0a}\ee

 The confining correlators $D^E, D^H$ generate  the nonzero values of CE and CM string tensions,
 \be \sigma^{E(H)} = \frac12 \int D^{E (H)} (z) d^2z.\label{1a}\ee

The CE correlators $D^E$ and $D^E_1$ produce   the scalar confining interaction
$V_D(R)$ and the vector-like nonperturbative  interaction $V_1(R)$
respectively.

\be V_D(R) = 2 c_a \int^r_0 (r-\lambda) d\lambda \int^\infty_0 d\nu D^E
(\lambda, \nu)= V_D^{(\rm lin)} (r)+ V^{(\rm sat)}_D (r) \label{3a}\ee
 \be V_1 (r)= c_a
\int^r_0 \lambda d\lambda \int^\infty_0 d\nu D_1^E (\lambda, \nu), ~~c_{\rm
fund} =1, ~~ c_{\rm adj} =9/4.\label{4a}\ee

  At
the beginning of nineties a new theory of temperature transition in QCD was
suggested in \cite{18, 19}, where at the critical temperature $T_c$ the
correlator $D^E$, and hence  CE confinement disappears, while the CM vacuum
fields survive.

The  advanced form of the  np theory of the thermalized QCD was given in
\cite{20}, where the Polyakov lines have been  derived from the vector CE
potential $V_1(r,T)$, produced by the CE field correlator $D^E_1(x-y)$.

Recently the approach of FCM for QCD at  $T>0$ was  reconsidered with the aim
to take into account the most important np contributions: vector CE interaction
$V_1 (r, T)$ at all $T$ and CM confinement at $T\geq T_c$.

It was shown in \cite{23}, that the latter phenomenon resolves the old Linde
problem, since it produces the effective CM Debye mass and eliminates IR
divergence  of perturbative theory, however justifying the necessity of summing
up the  infinite  series of diagrams in the order $O(g^6)$.

In \cite{25,27} the CM confinement  was taken into account together with exact
treatment of   Polyakov lines in the SU(3) theory. The resulting pressure
$P(T)$ and entropy $s(T)$ are in good agreement with lattice data \cite{40}.

It is  a purpose of the present paper to apply  the same method, as in
\cite{25,27}, to the analysis of the QCD matter with $n_f >0$ at $T\geq T_c$,
taking into account accurate values of Polyakov lines and the CM confinement.

 Below  we  explain the general formalism in section 2.

In section 3 the notation of the CM confinement and its dynamics is treated and
the resulting for formulas for $P(T)$, $s(T)$ are obtained. In section 4 the main
dynamical input is defined  with respect the $V_1 (r,T)$ and Polyakov lines
$L(T)$. In section 5 the numerical results are shown and discussed.

\section{General formalism}
In this section  we are  using thermodynamics of quarks and  gluons in the
vacuum background fields (VBF), as formulated in \cite{18}. For the gluon
contribution one obtains

\be
 \lan F_0^{gl}(B) \ran_B= -T\int^\infty_0\frac{ds}{s} \xi(s) d^4
 x(Dz)^w_{xx} e^{-K}\left [ \frac12 tr \lan \tilde
 \Phi_F(x,x)\ran_B-\lan tr \tilde \Phi (x,x)\ran \right].
 \label{1}
 \ee
Here $ B_\mu$ refers to the VBF, $K=\frac14 \int^s_0  d\tau\left( \frac{dz_\mu
(\tau)}{d\tau} \right)^2,$ $\xi(s)$ is a regularizing factor at $s=0$, and

 \be \tilde \Phi_F(x,y) = P_FP\exp(2ig \int^t_0 \hat F(z(\tau)) d\tau)\exp
(ig\int^x_y \tilde B_\mu dz_\mu).\label{2}\ee Here $P_F, P$ are  the ordering
operators,  and $\hat F$ is  the field strength of the field $B_\mu$, also
$\tilde \Phi(x,y)$  is obtained from (\ref{2}) with  $\hat F=0$. The winding
path measure is
\begin{eqnarray}
(Dz)^w_{xy}&&= \lim_{N\to \infty}\prod^N_{m=1}
\frac{d^4\zeta(m)}{(4\pi\varepsilon)^2} \nonumber
\\
&& \sum_{n=0,\pm,...} \frac{d^4p}{(2\pi)^4}\exp \left[ ip_\mu\left(
\sum^N_{m=1} \zeta (m)-(x-y)_\mu-n\beta \delta_{\mu 4}\right)\right]. \label{3}
\end{eqnarray}

As one  can see  in (\ref{1}), there enters the adjoint gluon loop  $tr \tilde
\Phi (x,x)$, which will be a major point of our investigation.

Using relation  $P_{gl} V_3 = - \lan F_0  (B) \ran_B$, one  can   rewrite
(\ref{1}) as \be P_{gl} = (N_c^2-1) \int^\infty_0 \frac{ds}{s} \sum_{n\neq 0 }
G^{(n)} (s),~~ G^{(n)} (s) = \int (Dz)^w_{on} e^{-K}  \hat{tr}_a\lan W_{\sum} (C_n)\ran.\label{4}\ee

Here $W_{\sum}$ is  the adjoint Wilson loop with  the   contour $C_n$, and
$\hat{tr}_a$ is  the normalized adjoint trace.

Note, that we have disregarded so far all perturbative contributions except
those possible inside the gluon loop.

We now turn to the quark contribution, which according to \cite{20}, can be
written in a form, similar to (\ref{4})

\be P_q = 2 N_c\int^\infty_0  \frac{ds}{s} e^{-m^2_q s} \sum^\infty_{n=1}
(-)^{n+1}[S^{(n)} (s) + S^{(-n)} (s)],\label{5}\ee  \be S^{(n)} (s) = \int (
{Dz} )^w_{on} e^{-K} \frac{1}{N_c}  \hat{tr } \lan W_\sigma
(C_n)\ran.\label{6}\ee

At this point it is important to look into  the details of the vacuum dynamics
at $T\geq T_c$, where the main contribution is given by the  correlators
$D_1^E$ and $D^H$. The first is  acting in the temporal surfaces $(i4)$, via
the interaction $V_1(r,T)$, which can be written, according to \cite{29*} as
\be V_1(r,T) = \int^\beta_0 d\nu (1-\nu T) \int^r_0 \xi d\xi D^E_1
(\sqrt{\xi^2+\nu^2}).\label{7}\ee

Separating, as in \cite{25, 27} the constant term $V_1(\infty, T)$, one obtains
in (\ref{4}), (\ref{6}) a factorization of the space 3d, where $D^H$ is acting,
and the temporal direction, which yields \cite{18, 25, 27} \be G^{(n)}(s) =
\int (Dz_4)^w_{on} e^{-K_4-J_n^E} G_3 (s),~~  S^{(n)}(s) = \int
(\overline{Dz_4})^w_{on} e^{-K_4-J_n^E} S_3 (s) .\label{8}\ee

Here  $G_3(s), S_3(s)$ are 3 d closed loop Green's  functions

 \be G_3(s)=
\int (D^3 z)_{xx} e^{-K_{3d}} \lan \hat{tr}_a W_3^a\ran,~~S_3(s)= \int (D
z)_{xx} e^{-K_{3d}} \lan \hat{tr}_f W_3^f\ran.\label{9}\ee

As shown in \cite{29*},  $V_1$ enters  in $J_n^E$, which  contributes to PL \be
J_n^E = \frac{n\beta}{2} \int^{n\beta}_{0} d\nu\left(
1-\frac{\nu}{n\beta}\right) \int^\infty_0 \xi d\xi D_1^E (\sqrt{\xi^2+\nu^2}),
~~  L_{\rm adj}^{(n)} =\exp \left( - \frac{9}{4} J^E_n\right).\label{10}\ee

One can see in (\ref{10}), that for $T\la  M_{g lp}=1.5$ GeV, (the gluelump
mass) and $n<n^*=
 \frac{M_{glp}}{T}$, $J^E_n\approx nJ^E_1$, and hence $  L_{\rm adj}^{(n)}
 \approx ( L_{\rm adj})^{ n }, ~~  L_{\rm adj}\equiv L_{\rm adj}^{(1)}$.

 Here $ {M_{glp}}^{-1}$ is the range of $D^E_1$, as was discussed in \cite{28}.

 Inserting over $(Dz_4)$ in (\ref{8}), one obtains  $\int (D z_4)^w_{on} e^{-K_4}
  = \frac{1}{\sqrt{4\pi s}} e^{-\frac{n^2\beta^2}{4s}}$
  and one has the following form  for gluon pressure \cite{25}

 \be P_{gl} = \frac{N_c^2-1}{\sqrt{4\pi}} \int^\infty_0 \frac{ds}{s^{3/2}}
G_3 (s) \sum_{n=\pm 1,\pm 2,..} e^{-\frac{n^2}{4T^2s} } L_{\rm
adj}^{(n)}.\label{11}\ee

In  a similar way  from (\ref{5})--(\ref{9}) one obtains the quark pressure for
one quark flavor with the mass $m_q$.

\be P_{q} = \frac{4N_c }{\sqrt{4\pi}} \int^\infty_0 \frac{ds}{s^{3/2}}
 e^{-m^2_qs}
S_3 (s) \sum_{n=  1,2,..}(-)^{ n+1 } e^{-\frac{n^2}{4T^2s} }
L_{f}^{(n)}.\label{12}\ee

In the next section we analyze the 3d loop  CM contributions in $S_3(s),
G_3(s)$.

\section{Colormagnetic confinement contribution to $S_3(s), G_3(s)$}

As one can see in (\ref{9}), $G_3(s)$ and $S_3(s)$  contain the contribution of
the adjoint and fundamental loops respectively, which are subject to the area
law, $\lan \hat{ tr}_i W_3\ran= \exp (-\sigma_i  {\rm are} a(W))$  $i= $ fund,
adj. Kinetic term is in $K_{3d}$ in (\ref{9}), so both $G_3(s)$ and $S_3(s)$
are proportional to the  Green's functions of two color charges, connected by
confining string, from one point $x$ on the loop to another (arbitrary) point,
e.g. the point  $u$ on the same loop.

There are two ways, how the CM confinement can be taken into account, suggested
in \cite{25}.  Considering the oscillator interaction between the charges, one
obtains \be G_3^{OSC} (s) = \frac{1}{(4\pi)^{3/2}\sqrt{s}} \frac{M_{\rm
adj}^2}{sh M_{\rm adj}^2 s}\label{17}\ee
 and $S^{OSC}_3 (s)$ is obtained from (\ref{17}), replacing $M_{\rm adj}$, by
 $M_f$. Here $M_{\rm adj} = 2 \sqrt{\sigma_s}= m_D (T)$, where $m_D (T)$ is the
 Debye mass, calculated in \cite{29} in good agreement with lattice data
 \cite{30}.

 A more realistic form obtains, when one replaces the linear interaction
 $\sigma_s r \to \frac{\sigma_s}{2} \left( \frac{r^2}{\gamma} + \gamma\right)$,
 varying the parameter $\gamma$ in the final expressions, imitating in this way
 linear interaction by an oscillator potential. Following \cite{25,27}   one obtains
 \be G_3^{\rm lin} (s) = \frac{1}{(4\pi s)^{3/2}} \left( \frac{ M^2_{\rm adj}
 s}{sh (M_{\rm adj}^2 s)}\right)^{1/2}, ~~ S_3^{\rm lin} (s) = G_3^{\rm lin}
 (s) |_{M_{\rm adj}\to M_f},\label{18}\ee
where we have taken into account as  in (\ref{9}), that $S_3$ obtains from
$G_3$ replacing adjoint loop $W_3$ by the fundamental one.
 Finally, substituting these expressions in (\ref{11}), (\ref{12}), one obtains
 the equations for $P_{gl}^{\rm lin}, P_q^{\rm lin}$, containing the effects of CM
 confinement, which  will be used in what follows.

\section{  $V_1(r,T)$ and the Polyakov lines}

In this section  we analyze Polyakov lines  (PL) $L_i , i=adj, f,$ and the $np$
interaction $V_1^{np}(r,T)$, which generates those as functions of temperature.
It is fundamentally important, that $V_1^{np}(r,T)$ has  a finite nonzero limit
at large $r$ , as it is seen in (\ref{7}), and it is  exactly this value which
enters  in  $L_i$ at not  large $n$, \be  L_i^{(n)} \simeq \exp \left( -c_i n
\frac{ V_1^{np} (\infty, T)}{2T} \right), ~~ c_{adj} = \frac94, ~~ c_f =1.
\label{19}\ee

On the lattice $PL$  can be measured in two ways, from the correlator of two
$PL$ at the distance $r$, which yields the  singlet free energy $F_{Q\bar Q}^s
(r,T)$ \cite{33}, which is equivalent to $V_1(r,T)$, and includes also the
perturbative contributions.

On the other hand, $F_q(T)$ can be found together with $L_f$ from the direct
measurement of the fundamental line \be L^{\rm bare} = \frac13 \left\lan Tr
\prod^{N_c-1}_{x_0 =0} U_0 (\vex, x_0)\right \ran_{\rm vac, \vex}.\label{22}\ee

The resulting values of the renormalized $L$ are strongly dependent on the type
of lattice quark operator used.

In what  follows we shall take our $L_f$ using  our $V_1^{np}$ from \cite{29*},
which are in agreement with data from \cite{34}. More explicitly we are writing
for $V_1$ as in \cite{20} \be V_1^{np} (\infty, T) = \frac{0.175~{\rm
GeV}}{1.35\frac{T}{T_c} -1}. \label{21a}\ee

  We
shall be  using these values of $V_1^{np}$ and the corresponding values of
$L_i(T)$, Eq. (\ref{19}),   in our Eqs. (\ref{11}), (\ref{12}), where $S_3$ and
$ G_3$ are given in (\ref{18}) to obtain $P(T), s(T)$ and compare  to lattice
data.

\section{Results and discussion}

In this section we present our results for $P(T)$, $s(T)$ in the  temperature
region 150~MeV~$\leq T \leq $~1000~MeV. For $P(T) =\sum_{m_q} P_q^{(m_q)} (T)
+P_g(T)$  we are using  Eqs. (\ref{11}), (\ref{12}) with $G_3(s), S_3(s)$ from
(\ref{18}). The Polyakov lines are obtained from  (\ref{19}), (\ref{21a}). We
are using $m_q =3,5$ and 100 MeV for $m_q = m_u, m_d$ and $m_s$ respectively.

We compare in Fig. 1  our results for $\frac{P(T)}{T^4}$  with the lattice
results from Table 3 (right column) of \cite{40}. In the following Fig.2  we
report our results for  $s(T)$ in comparison with lattice data from \cite{40}.

{
\begin{figure}[htb] 
\setlength{\unitlength}{1.0cm}
\centering
\begin{picture}(9.0,6.0)
\put(0.6,0.6){\includegraphics[height=5.4cm]{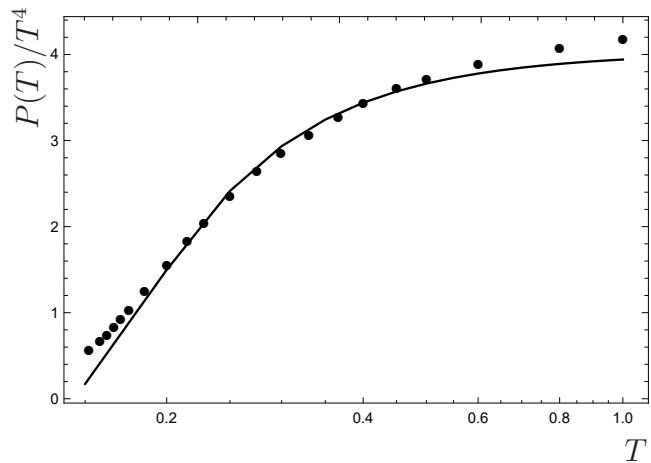}}
\put(8.15,0.1){$T$}
\put(0.1,4.5){\rotatebox{90}{$P(T)/T^4$}}
\end{picture}
\vspace{-0.15cm}
\caption{The pressure $\frac{P(T)}{T^4}$, as obtained from Eqs. (\ref{11}), (\ref{12}), -- solid line, in comparison with lattice data from \cite{40} -- filled dots.}
\label{fig:fig01}
\end{figure}
} 

{
\begin{figure}[htb] 
\setlength{\unitlength}{1.0cm}
\centering
\begin{picture}(9.0,6.0)
\put(0.6,0.6){\includegraphics[height=5.4cm]{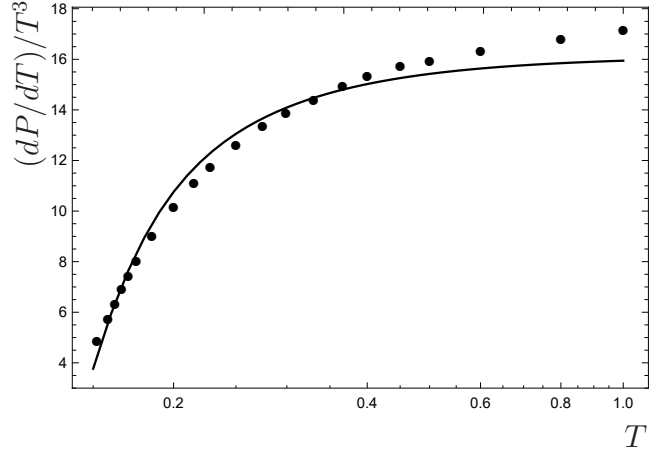}}
\put(8.15,0.1){$T$}
\put(0.1,3.8){\rotatebox{90}{$(dP/dT)/T^3$}}
\end{picture}
\vspace{-0.15cm}
\caption{$\frac{s(T)}{T^3}=\frac{1}{T^3}\frac{dP(T)}{dT}$ for the pressure given in Fig. \ref{fig:fig01} -- solid line, and the lattice data from \cite{40} -- filled dots.}
\label{fig:fig02}
\end{figure}
} 

One can see in both Figs.1 and 2 a good agreement of our results with  lattice
data. Comparing this with our $P(T)$  from \cite{20}, where the same $V_1^{np}$
 and Polyakov loops  were exploited, but CM confinement in $S_3, G_3$ was absent,
one can deduce, that the CM contribution is very important in the whole
interval of $T$ up to 1 GeV. The same is true also for the pure $SU(3)$ theory,
studied in \cite{25,27}. Moreover, in \cite{23} it was shown, that CM
confinement solves the old Linde problems, preventing the accurate perturbative
calculations in the region $T<600$ MeV.

Our results show that the FC method can be successfully applied to the
quark-gluon thermodynamics and in  particular it is planned to extend our
analysis to the case of nonzero chemical potential.

We specifically excluded from our analysis
the region $T<150$ MeV, where the correlator $D^E(x)$ is acting, since the interesting mechanisms of deconfinement and mutual replacements of $V_D$ and $V_1^{np}$ in this region, discussed in \cite{27},
require more space and planned for the future.

The authors are grateful for useful discussions to
B.O.~Kerbikov and M.A.~Andreichikov.

This work was done in the framework of  the scientific  project,
   supported by the
     Russian Science Foundation grant   number 16-12-10414.

 \end{document}